\documentclass[sn-nature]{sn-jnl}

\usepackage{graphicx}%
\usepackage{multirow}%
\usepackage{amsmath,amssymb,amsfonts}%
\usepackage{amsthm}%
\usepackage{mathrsfs}%
\usepackage[title]{appendix}%
\usepackage{xcolor}%
\usepackage{textcomp}%
\usepackage{manyfoot}%
\usepackage{booktabs}%
\usepackage{algorithm}%
\usepackage{algorithmicx}%
\usepackage{algpseudocode}%
\usepackage{listings}%
\usepackage{mhchem}%
\usepackage{xcolor}

\theoremstyle{thmstyleone}%
\theoremstyle{thmstyletwo}%

\theoremstyle{thmstylethree}%

\begin{document}

\title[Title]{Clarifying the Radiative Decay of the Hoyle State with Charged-Particle Spectroscopy}


\author*[1,2]{\fnm{D.} \sur{Dell'Aquila}}\email{daniele.dellaquila@unina.it}

\author*[3,4]{\fnm{I.} \sur{Lombardo}}\email{ivano.lombardo@ct.infn.it}

\author[3,4]{\fnm{L.} \sur{Redigolo}}
\author[1,2]{\fnm{M.} \sur{Vigilante}}

\author[6,7]{\fnm{F.} \sur{Angelini}}
\author[8,9]{\fnm{L.} \sur{Baldesi}}
\author[8,9]{\fnm{S.} \sur{Barlini}}
\author[1,2]{\fnm{A.} \sur{Best}}
\author[8,9]{\fnm{A.} \sur{Camaiani}}
\author[9]{\fnm{G.} \sur{Casini}}
\author[8,9]{\fnm{C.} \sur{Ciampi}}
\author[6,7]{\fnm{M.} \sur{Cicerchia}}
\author[4]{\fnm{M.} \sur{D'Andrea}}
\author[11]{\fnm{J.} \sur{Dikli\'c}}
\author[7]{\fnm{D.} \sur{Fabris}}
\author[5]{\fnm{B.} \sur{Gongora Servin}}
\author[5]{\fnm{A.} \sur{Gottardo}}
\author[5]{\fnm{F.} \sur{Gramegna}}
\author[1,2]{\fnm{G.} \sur{Imbriani}}
\author[5]{\fnm{T.} \sur{Marchi}}
\author[10]{\fnm{A.} \sur{Massara}}
\author[6,7]{\fnm{D.} \sur{Mengoni}}
\author[2]{\fnm{A.} \sur{Ordine}}
\author[11]{\fnm{L.} \sur{Palada}}
\author[8,9]{\fnm{G.} \sur{Pasquali}}
\author[9]{\fnm{S.} \sur{Piantelli}}
\author[6,7]{\fnm{E.} \sur{Pilotto}}
\author[1,2]{\fnm{D.} \sur{Rapagnani}}
\author[11]{\fnm{M.} \sur{Sigmund}}
\author[8,9]{\fnm{A.} \sur{Stefanini}}
\author[6,7]{\fnm{D.} \sur{Stramaccioni}}
\author[12]{\fnm{D.} \sur{Tagnani}}
\author[11]{\fnm{I.} \sur{Ti\v{s}ma}}
\author[9]{\fnm{S.} \sur{Valdr\'e}}
\author[4]{\fnm{G.} \sur{Verde}}
\author[11,13]{\fnm{N.} \sur{Vukman}}

\affil[1]{\orgdiv{Dipartimento di Fisica "Ettore Pancini"}, \orgname{University of Naples "Federico II"}, \orgaddress{\city{Naples}, \country{Italy}}}

\affil[2]{\orgdiv{Sezione di Napoli}, \orgname{INFN}, \orgaddress{\city{Naples}, \country{Italy}}}

\affil[3]{\orgdiv{Dipartimento di Fisica "Ettore Majorana"}, \orgname{University of Catania}, \orgaddress{\city{Catania}, \country{Italy}}}

\affil[4]{\orgdiv{Sezione di Catania}, \orgname{INFN}, \orgaddress{\city{Catania}, \country{Italy}}}

\affil[5]{\orgdiv{Laboratori Nazionali di Legnaro}, \orgname{INFN}, \orgaddress{\city{Legnaro}, \country{Italy}}}

\affil[6]{\orgdiv{Dipartimento di Fisica}, \orgname{University of Padova}, \orgaddress{\city{Padova}, \country{Italy}}}

\affil[7]{\orgdiv{Sezione di Padova}, \orgname{INFN}, \orgaddress{\city{Padova}, \country{Italy}}}

\affil[8]{\orgdiv{Dipartimento di Fisica e Astronomia}, \orgname{University of Firenze}, \orgaddress{\city{Sesto Fiorentino}, \country{Italy}}}

\affil[9]{\orgdiv{Sezione di Firenze}, \orgname{INFN}, \orgaddress{\city{Sesto Fiorentino}, \country{Italy}}}

\affil[10]{\orgdiv{Laboratori Nazionali del Sud}, \orgname{INFN}, \orgaddress{\city{Catania}, \country{Italy}}}

\affil[11]{\orgdiv{Zavod za eksperimentalnu fiziku}, \orgname{Rudjer Bo\v{s}kovi\'c Institute}, \orgaddress{\city{Zagreb}, \country{Croatia}}}

\affil[12]{\orgdiv{Sezione di Roma Tre}, \orgname{INFN}, \orgaddress{\city{Rome}, \country{Italy}}}

\affil[13]{Currently also with \orgdiv{Sezione di Perugia}, \orgname{INFN}, \orgaddress{\city{Perugia}, \country{Italy}}}

\abstract{A detailed knowledge of the decay properties of the so called Hoyle state in the $^{12}$C nucleus ($E_x=7.654$ MeV, $0^+$) is required to calculate the rate at which carbon is forged in typical red-giant stars. This paper reports on a new almost background-free measurement of the radiative decay branching ratio of the Hoyle state using advanced charged particle coincidence techniques. The exploitation, for the first time in a similar experiment, of a bidimensional map of the coincidence efficiency allows to reach an unitary value and, consequently, to strongly reduce sources of systematic uncertainties. The present results suggest a value of the radiative branching ratio of $\Gamma_{rad}/\Gamma_{tot}=4.2(6)\cdot10^{-4}$. This finding helps to resolve the tension between recent data published in the literature.}

\maketitle

Carbon is a fundamental chemical constituent for the existence of life on Earth. It is mostly forged in stars in the so called \emph{triple-alpha process}, a nuclear reaction that turns three $\alpha$-particles (nuclei of $^4$He) into a $^{12}$C nucleus \cite{Clayton}. At temperatures typical of red-giant stars, this process proceeds through a peculiar excited state of the $^{12}$C nucleus named the Hoyle state ($E_x=7.654$ MeV, $0^+$) \cite{Freer14}. The rate of carbon formation is thus determined by the nuclear properties of the Hoyle state, namely by the competition of its very rare \emph{radiative decay} and its dominant particle-decay. However, while there is a general consensus regarding the latter \cite{DellAquila17,Smith17,Rana19,Bishop20,Smith20}, there is presently tension in the determination of the radiative decay \emph{branching ratio}. A recent experiment \cite{Kibedi20} suggested a radiative branching ratio larger by a factor $1.5$ than the commonly accepted value derived from elder experiments \cite{Alburger61,Seeger63,Hall64,Chamberlin74,Davids75,Mak75,Markham76,Obst76}. This would result in an increase of about $34 \%$ of the triple-alpha reaction rate, directly affect the carbon-to-oxygen balance in the Universe, and have a dramatic impact on the pair-instability mass gap for the formation of black holes \cite{Woosley21}. This newly suggested value has been subsequently contradicted by two additional experiments \cite{Tsumura21,Luo24} (the paper of Ref.~\cite{Luo24} was published during the review process of the present manuscript.); at variance, the only existing experiment detecting all the reaction products \cite{Cardella21} reports a value even larger than \cite{Kibedi20}, but affected by a large uncertainty. This tension clearly calls for a new, independent, probe of the radiative branching ratio of the Hoyle state. In this work, we report on an almost background-free measurement of the radiative decay branching ratio of the Hoyle state using charged particle coincidence techniques. We exploit a deuteron on nitrogen nuclear reaction to produce, in a terrestrial laboratory, $^{12}$C nuclei excited in the Hoyle state. The radiative branching ratio is directly deduced by counting their total number and the number of events in which they lead, after radiative emission, to a $^{12}$C in its ground state. The present particle-particle coincidence experiment adopted several techniques to minimize the background and clearly identify the signal associated to the radiative decay. Furthermore, for the first time in a charged particle coincidence experiment, we performed a careful topological study of the coincidence detection efficiency in two dimensions, allowing to reach unitary coincidence efficiency and to have under full control one of the most critical sources of systematic uncertainties. The new findings contradict the result of \cite{Kibedi20} and are in excellent agreement with the former estimate of the literature \cite{Freer14}.

\begin{figure}[]%
	\centering
	\includegraphics[width=0.8\textwidth]{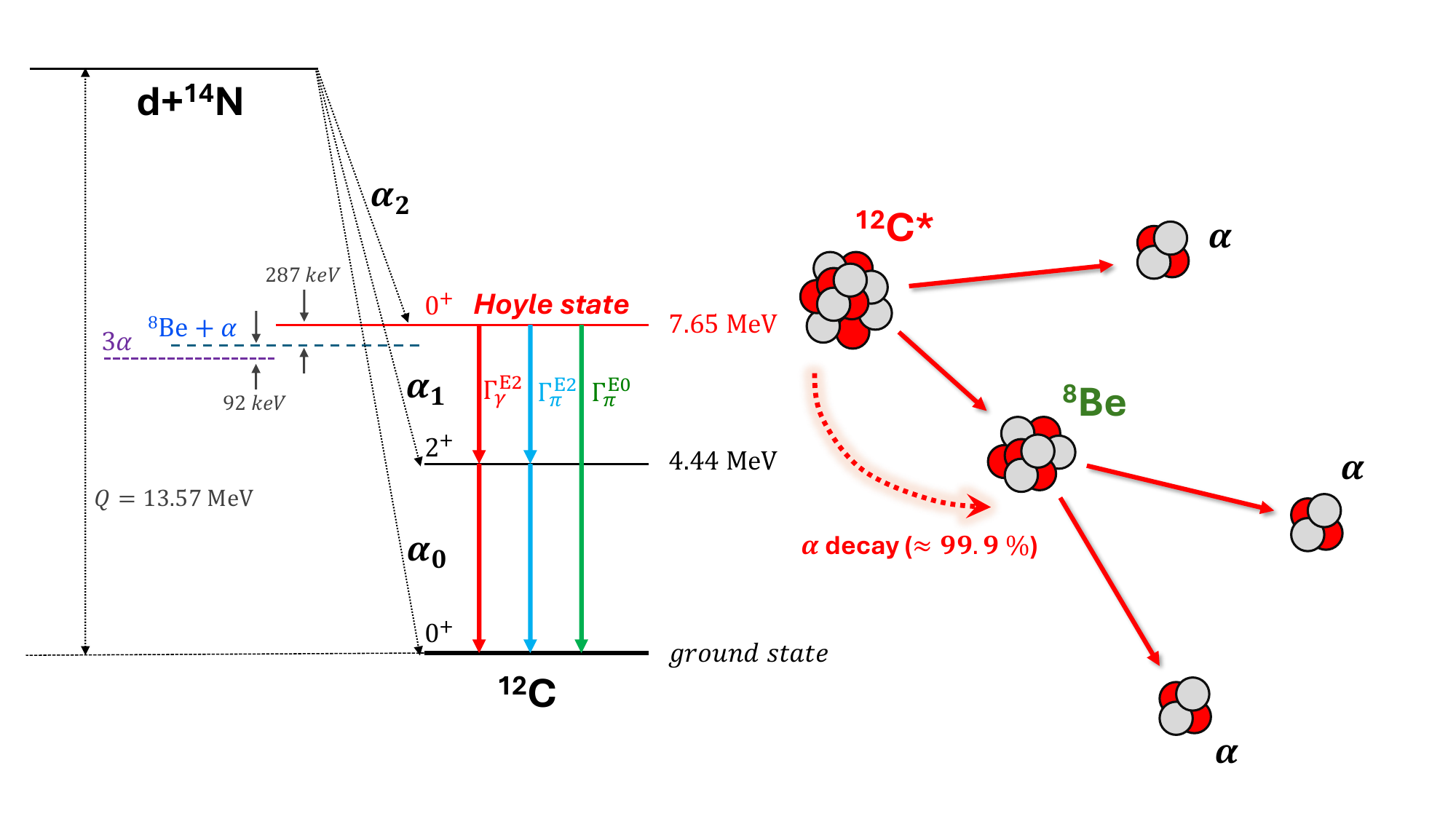}
	\caption{A schematic diagram of the first low-lying states of $\ce{^{12}C}$ and the decay modes of the Hoyle state. The main $\ce{d + ^{14}N}$ reactions populating the $\ce{^{12}C}$ ground state ($\alpha_0$), the $4.44$ MeV state ($\alpha_1$), and the Hoyle state ($\alpha_2)$ are indicated. The various radiative decay modes of the Hoyle state (a cascade of two $\gamma$-rays, predominant, and pair-emission decays) are outlined. The $\alpha$ decay is schematically depicted assuming a sequential mechanisms through $\ce{^{8}Be}$.}
	\label{fig:level_scheme}
\end{figure}
\begin{figure}[]%
	\centering
	\includegraphics[width=0.8\textwidth]{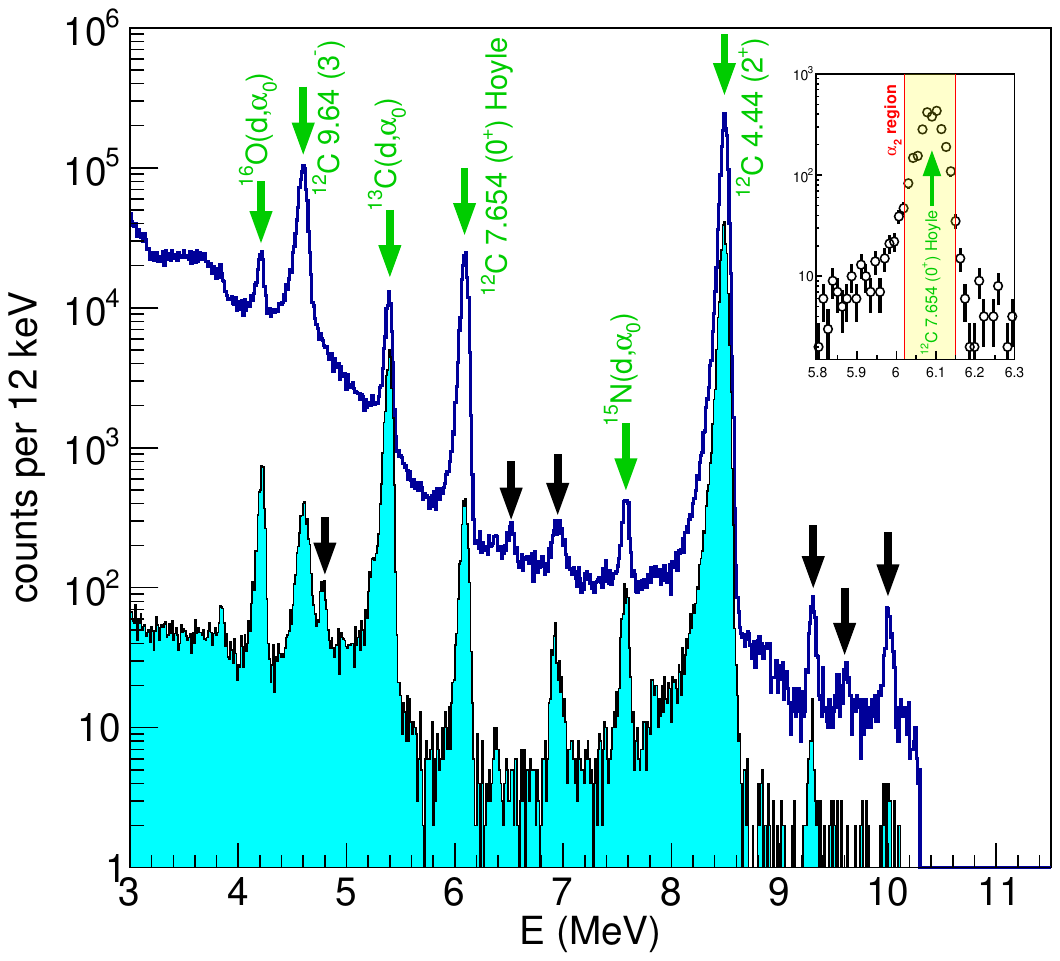}
	\caption{(Blue spectrum) Energy spectrum obtained with the backward telescope, used as anti-coincidence telescope. Green arrows indicate ejectile energies calculated for the main expected reactions. Black arrows are the energy positions of reactions occurring on minor contaminants present in the target, such as $\ce{^7Li}$, $\ce{^9Be}$, $\ce{^{19}F}$. (Cyan-filled spectrum) Same as the blue spectrum, but obtained using only events in which a coincident hit is present at the forward telescope. The background in the proximity of the Hoyle peak is strongly suppressed in the coincidence data. (Insert) A zoomed view of the cyan-filled spectrum in proximity of the $\alpha_2$ peak.}
	\label{fig:dietro}
\end{figure}
\begin{figure}[]%
	\centering
	\includegraphics[width=0.35\textwidth]{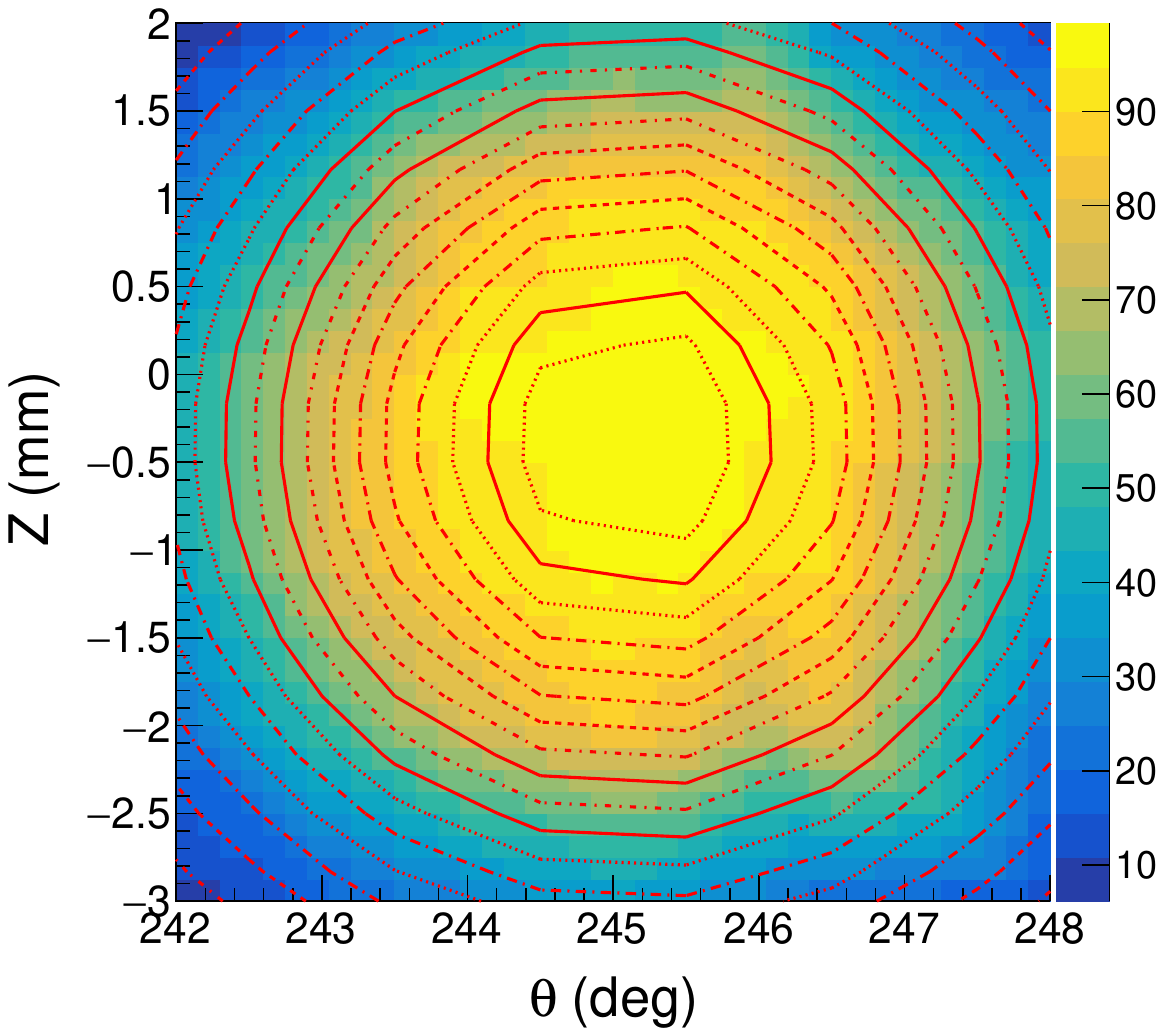}
	\includegraphics[width=0.6\textwidth]{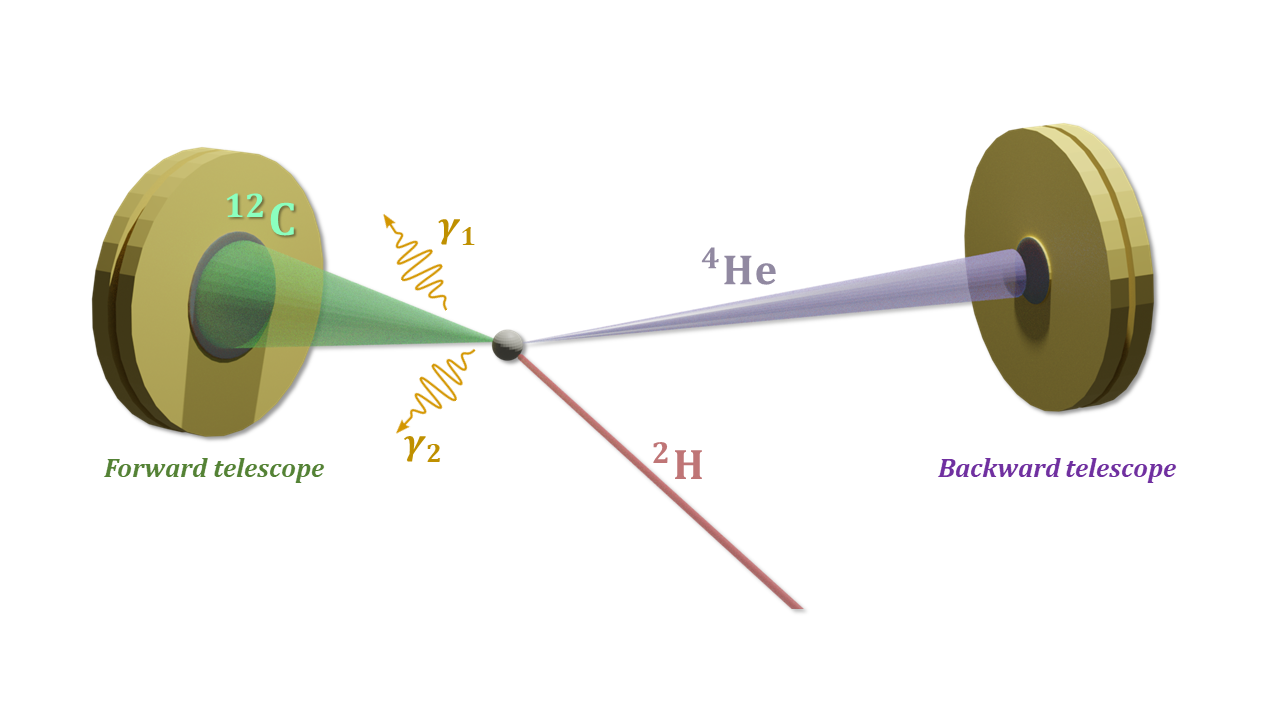}
	\caption{(left) Bi-dimensional $z$-$\theta$ efficiency map for $\alpha_1$-$\ce{^{12}C}$ coincidence events. Values of $\theta$ are expressed in an arbitrary reaction chamber reference frame. Unitary coincidence efficiency is safely achieved in a broad $z$ and $\theta$ domain. This plot is used to center the forward telescope with the $\ce{^{12}C^\star}$ ($4.44$ MeV) kinematic emission cone. (right) Schematic layout of the experiment. In their nominal positions, the backward telescope is placed at $\theta=90^\circ$, while the forward telescope is placed at $\theta=64.8^\circ$ with respect to the $\ce{^{2}H}$ beam direction.}
	\label{fig:efficiency}
\end{figure}

We exploited the $\ce{^{14}N}$(d,$\alpha_2$)$\ce{^{12}C^\star}$ nuclear reaction to produce excited $\ce{^{12}C}$ nuclei; the subscript $_2$ indicating the emission of $\ce{^{12}C}$ in its second excited state (i.e. the Hoyle state, see Fig.~\ref{fig:level_scheme}). At low bombarding energies, this reaction is particularly selective to low-lying $^{12}$C excited states, as discussed, for example, in Ref.~\cite{DellAquila17}. A charged particle detector (\textit{backward}, hereafter) was used to tag the excitation energy of the $\ce{^{12}C}^\star$ recoil by measuring the energy of the emitted $\alpha$ particle, as shown by the blue spectrum of Fig.~\ref{fig:dietro}. The events in which the recoiling $\ce{^{12}C^\star}$ is populated in its Hoyle state are selected focusing on a narrow energy region around the expected $\alpha_2$ energy. Due to the two-body nature of the reaction, this poses a stringent constraint on the momentum vector of the associated $\ce{^{12}C^\star}$. Another charged particle detector (\textit{forward}, hereafter) is accurately placed to measure, with unitary coincidence efficiency and almost-zero background, the coincident recoil after its radiative decay to the ground state (Fig.~\ref{fig:efficiency}, right). The radiative branching ratio is obtained as the ratio of the $\alpha_2$-$\ce{^{12}C}$ coincidence rate between the two detectors and the $\alpha_2$ rate measured by the backward detector without the coincidence condition.

Each detector was a telescope made by two longitudinally-stacked semiconductor detection units. The anti-coincidence technique, which utilizes a logic condition between the signals produced by the two detection stages of a given telescope, was exploited both in the backward and forward telescopes to strongly reduce the background due to other contaminant reactions \cite{DellAquila17}. The experiment used a highly collimated deuteron beam at $2.7$ MeV, produced at the CN accelerator of INFN Laboratori Nazionali di Legnaro, at typical intensities of $10$-$40$ nA. The beam intensity was chosen suitably low to make negligible the probability of spurious coincidences between the two telescopes. During the experiment, we measured a detection rate at the forward detector $<1000$ Hz, while the $\alpha_2$ rate at the backward detector was $<0.1$ Hz. Given that the time coincidence window was $\approx5$ ns, one expects a spurious coincidence rate $<10^{-6}$ Hz. In addition, it is worth noting that the most probable contaminants are elastically scattered deuterons and protons from (d,p) reactions, whose energy depositions in the forward detector (lower than $\approx1.3$ MeV) would be much lower than the energy expected for the $\ce{^{12}C}$ recoils. In order to ensure the reproducibility of the beam position on target and a good definition of the reaction vertex, the beam passed through a series of six circular collimators, with diameters ranging from $1$ mm to $3$ mm, placed before and after the reaction chamber. To this end, during the whole experiment, we constantly monitored the beam current on the collimators and on the Faraday cup, after the beam passes through the entire set of collimators. A similar technique to constrain beam emittance on target is used, for example, in Ref.~\cite{Mak75}. The resulting beam spot on the target had a diameter smaller than 1 mm. Nuclear reactions were induced on a thin melamine ($\ce{C_3H_6N_6}$) target ($\approx 50$ $\mu$g/cm$^2$ thick), deposited on an ultra-thin carbon backing ($\approx 10$ $\mu$g/cm$^2$). A tiny gold deposition ($\approx1.5$ nm) on the melamine helped to maintain the target stability under bombardment. The backward telescope was placed at a fixed position with its center at $\theta=90^\circ$ with respect to the beamline, while the forward telescope (a $17$ $\mu$m/$500$ $\mu$m silicon telescope) was mounted on a movable plate located in the forward hemisphere, which could be rotated about the experiment target. The vertical coordinate of the backward telescope was remotely controlled through a micro-positioning device with an accuracy better than $1$ $\mu$m.

A crucial point of the experiment is the capability to reconstruct the $\alpha_2$-$\ce{^{12}C}$ coincidence rate with unitary efficiency. To this end, the relative position between the backward and forward telescopes was accurately calibrated. Initially, to estimate the required forward detector solid angle when the backward detector is placed in a given position and with a given aperture, we carried out detailed Monte Carlo simulations of the $\ce{^{14}N}$(d,$\alpha_2$)$\ce{^{12}C^\star}$ reaction followed by the radiative decay of the Hoyle state. These simulations fully accounted for all the effects of interaction of the radiation with the matter and the beam optics. Detectors were then conservatively placed in their \emph{nominal} positions to fulfill the geometrical requirements suggested by the simulations: the forward detector covering a solid angle well larger (by about $25\%$) than the one needed to completely detect the $\ce{^{12}C}$ cone produced in the $\alpha_2$-$\ce{^{12}C}$ coincidence. Furthermore, the relative angle between the two detectors was carefully calibrated measuring the coincidence rate at the two detectors when the first $\ce{^{12}C}$ excited state ($4.44$ MeV, $2^+$) is populated and the forward detector is positioned around the expected direction of the $\ce{^{12}C^\star}$ recoil (i.e. $\theta=68.3^\circ$). The corresponding $\alpha_1$ particle detected by the backward detector is tagged restricting to events around the $\alpha_1$ energy in the plot of Fig.~\ref{fig:dietro}. Because the first excited state of $^{12}$C decays exclusively, i.e. with unitary branching ratio, to the ground state via a radiative transition, the $\alpha_1$-$\ce{^{12}C}$ rate must be equal to the $\alpha_1$ rate. In Fig.~\ref{fig:efficiency}, we show the ratio between these two rates as a function of the angular position of the forward detector (varied with $0.1^\circ$ precision using the movable plate) and its vertical coordinate (varied using the micro-positioning device). This plot was then used as a bi-dimensional coincidence efficiency map. A large spot where the efficiency is unitary is clearly identified, due to the conservative detector placement. The efficiency peak had a FWHM of $4.21$ mm and $5.09^\circ$, respectively for the vertical and horizontal axes. These values are well larger compared to the accuracy of the $Z$ and $\theta$ placement reached in the experiment, confirming the reliability of the unitary efficiency assumption. The small vertical displacement of the detector compared to its nominal position, which was previously accurately determined with the help of a laser level, is found to amount at $-0.36\pm0.04$ mm and was carefully compensated during the experiment. Once the relative position of the two telescopes is set-up to safely make the efficiency unitary, the forward detector is rigidly rotated by $\Delta \theta=3.5^\circ$ towards the beam direction to be centered to the angle expected for the $\alpha_2$-coincident $^{12}$C; the $\Delta\theta$ of this rotation is distinctively determined via kinematical considerations. The upper limit of possible out-of-plane deviations was verified exploiting a 3D reconstruction of the reaction chamber obtained with a laser scanner and found to be smaller than one hundredth of a degree, without effects on the efficiency. Possible efficiency variations due to small changes of the beam emittance on target were constantly monitored during the experiment by looking at the variations of the $\alpha_1$-$\ce{^{12}C}$ coincidence rate measured in the $\alpha_2$-$\ce{^{12}C}$ coincidence position. In addition, the $(\theta,Z)$ center of the bi-dimensional efficiency map was checked every $24$ hours of beam time. In general, possible systematic uncertainties on the overall efficiency of the coincidence measurement are estimated to be smaller than $2 \%$. This bi-dimensional search for unitary efficiency is done for the first time in a charged particle coincidence experiment, and is key to unambiguously probe the radiative branching ratio.

\begin{figure}[]%
	\centering
	\includegraphics[width=0.8\textwidth]{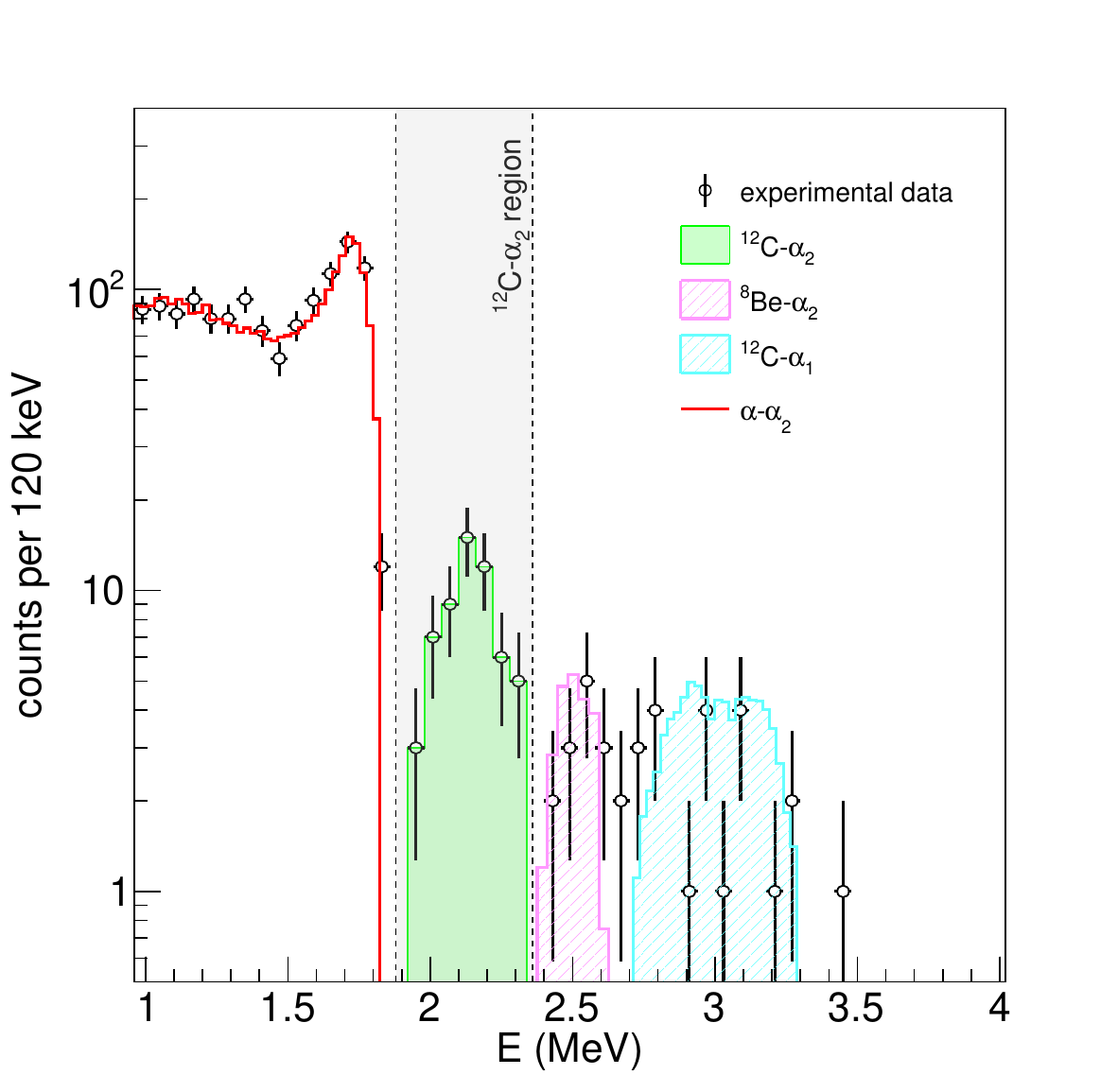}
	\caption{Forward energy spectrum in time-coincidence with the narrow energy region around the $\alpha_2$ peak shown in Fig.~\ref{fig:dietro}. The $\alpha_2$-$\ce{^{12}C}$ energy region is indicated.}
	\label{fig:davanti}
\end{figure}

To determine the radiative decay of the Hoyle state, the data analysis is restricted to $\alpha_2$ events selected in the plot of Fig.~\ref{fig:dietro}. After requiring that a valid hit is present in the forward detector in time-coincidence with the backward telescope, the blue spectrum of Fig.~\ref{fig:dietro} collapses to the cyan-filled one, where the continuous background, due to contaminant reactions, is strongly reduced. For time-coincident events in the proximity of the $\alpha_2$ energy, the rate of $\alpha_2$-$\ce{^{12}C}$ coincidence is obtained by analyzing the corresponding forward energy spectrum of Fig.~\ref{fig:davanti}. This spectrum exhibits four components. At low energy, a continuum due to $\alpha_2$-$\alpha$ coincidences is visible, well-described by the Monte Carlo simulation (red line). This contribution to the forward spectrum is dominant because the Hoyle state decays predominantly through $\alpha$-emission. It is characterized by a sharp fall, due to kinematic limits, and does not contaminate the expected $\alpha_2$-$\ce{^{12}C}$ energy region, which lies at higher energies, as shown in Fig.~\ref{fig:davanti}. $\alpha$-decay events lying at higher energies are possible only if a pileup of two-$\alpha$ particles, generated by the decay of $\ce{^{8}Be}$, occurs. However, a suitable choice of the target thickness and the gold layer deposition allows to boost these events outside of the $\alpha_2$-$\ce{^{12}C}$ region. This is possible thanks to the much larger energy deposition in the target matter of low-energy $^{12}$C ions compared to $\alpha$ particles, which results in enhancing their energy difference. The purple line shows the expected $\alpha_2$-$\ce{^{8}Be}$ contribution, simulated through Monte Carlo techniques. This is compatible, both for energy position and yield, to the observed experimental counts around $2.5$ MeV. At higher energies, one clearly observes contaminants due to a small fraction of $\alpha_1$-$\ce{^{12}C}$ events that fall in the detection cone of the forward detector, as suggested by the corresponding simulation (cyan line).

\begin{figure}[]%
	\centering
	\includegraphics[width=0.8\textwidth]{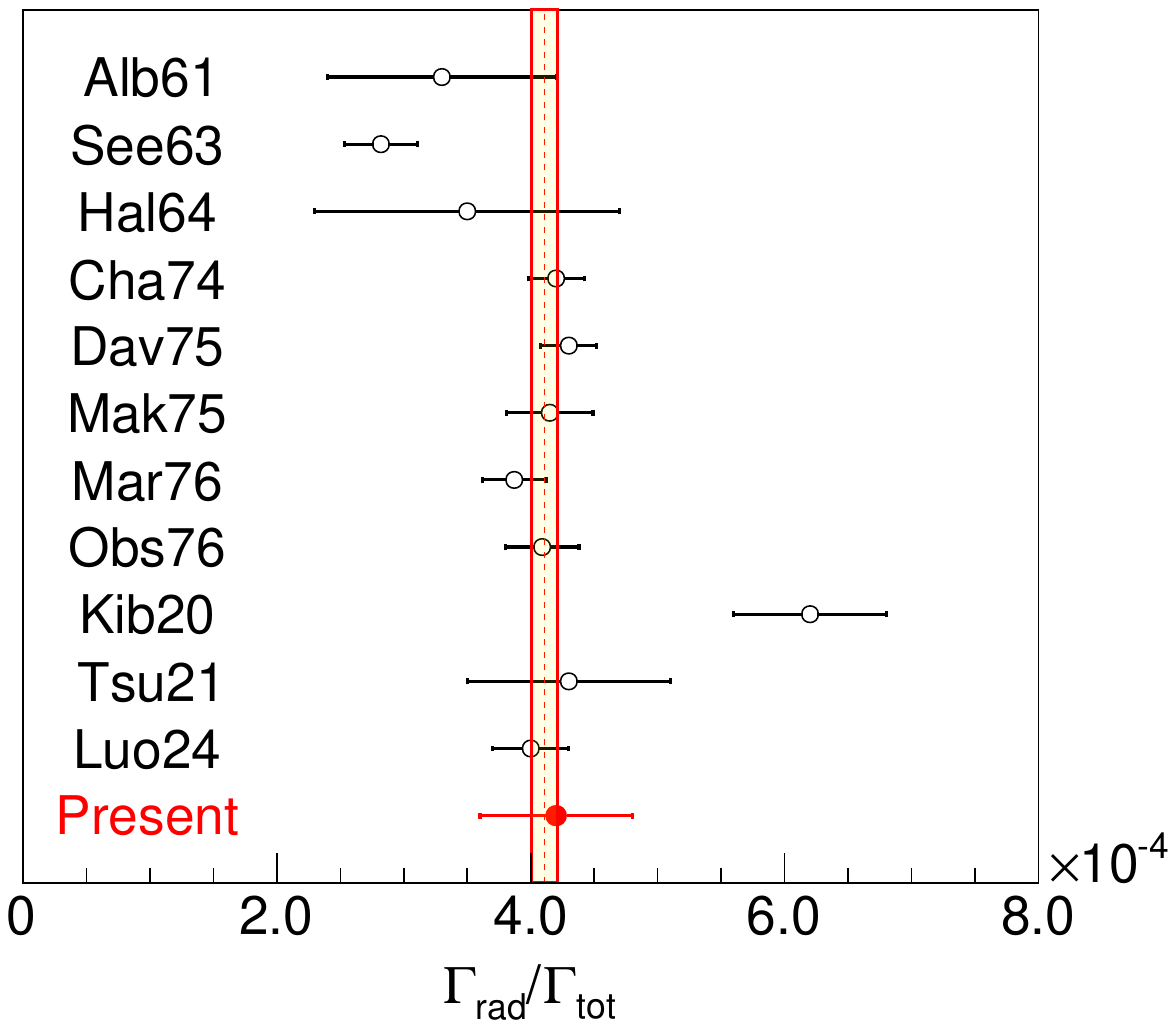}
	\caption{Radiative decay partial width $\Gamma_{rad}/\Gamma_{tot}$ of the Hoyle state in $^{12}$C from charged-particle coincidence experiments \cite{Luo24} (Luo24), \cite{Tsumura21} (Tsu21), \cite{Markham76} (Mar76), \cite{Mak75} (Mak75), \cite{Davids75} (Dav75), \cite{Chamberlin74} (Cha74), \cite{Hall64} (Hal64), \cite{Seeger63} (See63), and $\gamma$-particle coincidence experiments \cite{Kibedi20} (Kib20), \cite{Obst76} (Obs76), \cite{Alburger61} (Alb61). The result of \cite{Cardella21} ($\Gamma_{2\gamma}/\Gamma_{tot}=1.8\pm0.6\cdot10^{-3}$) is not shown for clarity reasons. The result of Ref.~\cite{Luo24} is also affected by a systematic uncertainty of $\pm0.16\cdot10^{-4}$, not shown in the figure. The present estimate is also shown. The yellow band indicates the new confidence interval obtained through a weighted average of all available estimates, excluding the values from \cite{Seeger63}, \cite{Kibedi20}, and \cite{Cardella21}.}
	\label{fig:hoyle_width}
\end{figure}

The situation illustrated in Fig.~\ref{fig:davanti} clearly allows us to deduce the $\alpha_2$-$\ce{^{12}C}$ rate with a simple energy selection. After background subtraction, we obtained $125190$ $\alpha_2$ counts tagged by the backward telescope without coincidence. The overall background amounted at about $2500$ counts. The systematic uncertainty on the estimation of the $\alpha_2$ event rate can be thus considered negligible compared to that arising from the efficiency estimation. Concerning the coincidence rate, in the region of interest for $\alpha_2$-$\ce{^{12}C}$ coincidences in the forward telescope, we observe $57$ counts. The background contribution ($4$ counts) was estimated by gating on nearby regions around the $\alpha_2$ peak of Fig.~\ref{fig:dietro}, as previously done in Ref.~\cite{Mak75}. Subtracting this small background, the obtained radiative decay branching ratio turns out to be $\Gamma_{rad}/\Gamma_{tot}=4.2(6)\cdot10^{-4}$. This value is in agreement with the presently accepted $\Gamma_{rad}/\Gamma_{tot}$ estimate (the recommended value from Ref.~\cite{Freer14}, based on the data from Refs. \cite{Alburger61,Hall64,Chamberlin74,Davids75,Mak75,Markham76,Obst76}, is $\Gamma_{rad}/\Gamma_{tot}=4.19(11)\cdot10^{-4}$). The value recently proposed in \cite{Kibedi20} ($6.2(6)\cdot10^{-4}$) is more than $3\sigma$ higher the present measurement: we suggest, therefore, to exclude it from the weighted average procedures for the determination of the Hoyle state radiative decay branching ratio. 

Fig.~\ref{fig:hoyle_width} summarizes $\Gamma_{rad}/\Gamma_{tot}$ values published in the literature up to date. Excluding the outlier values of Refs.~\cite{Seeger63,Kibedi20,Cardella21}, from the weighted average we recommend a radiative branching ratio $\Gamma_{rad}/\Gamma_{tot}=4.11(10)\cdot10^{-4}$.

The new findings do not support any significant revision of the triple-alpha reaction rate with respect to the traditional Caughlan and Fowler value \cite{Caughlan1988}. This result has crucial implications in astrophysics. A modification of the rate of the triple-alpha process has been in fact proposed as one of the possible solutions to the problem of the pair-instability mass gap for black holes: the mass limits have indeed being challenged by recent LIGO-Virgo observations on intermediate mass black holes \cite{Abbott20,Abbott20b}. Therefore, solutions to the puzzle on the pair-instability gap boundaries should be explored \textit{outside} the nuclear structure domain. Finally, after correctly accounting for the mechanisms of convection in stars \cite{Imbriani01,Straniero03}, clarifying the triple-alpha reaction rate poses tighter constraints on the rate of the "holy grail" $\ce{^{12}C}$($\alpha$,$\gamma$)$\ce{^{16}O}$ reaction \cite{Austin14}, as derived from observational data coupled to nucleosynthesis calculations in massive stars. The combined rates of these two reactions are key to explain the C/O abundance ratio in the Universe \cite{Clayton}.

\subsection*{Acknowledgements}
We gratefully acknowledge all the services of the research and accelerator divisions of INFN Laboratori Nazionali di Legnaro (Italy) for their valuable support in the preparation of the experiment. In particular, we are indebted to L. Maran, D. Lideo, L. La Torre, and D. Carlucci (INFN-LNL). We thank A. Barbon (University of Catania) for her support during the experimental data taking. We acknowledge the Target Laboratories of INFN Laboratori Nazionali del Sud (Catania, Italy) and INFN-LNL (M. Loriggiola and S. Carturan) for the development and analysis of the targets used in the present experiment. We thank V. Branchina (University of Catania), G. Cardella (INFN-Catania), and E. Perillo (University of Naples "Federico II") for useful discussions on the subject of this work. Finally, we express our gratitude to C. Rapicavoli (TMA, Catania) for his continuous and professional assistance in the construction of the high-precision mechanics needed for the experimental set-up. J.D., L.P., M.S., I.T., and N.V. acknowledge support of the Croatian Science Foundation under Project no. IP-2018-01-1257 and support of the Center of Excellence for Advanced Materials and Sensing Devices (Grant No. KK.01.1.1.01.0001). This work was partially supported by the Italian Ministry of Education, University and Research (MIUR) (PRIN2022 CaBS, CUP:E53D230023).

\subsection*{Data availability}
All relevant data used to produce these results are included in the manuscript.








\bibliography{bibliography_20230313}

\end{document}